\documentclass[referee]{raa_twocolumn}           
\usepackage{graphicx,times}
\usepackage{natbib}
\usepackage{amssymb,amsmath}
\bibpunct{(}{)}{;}{a}{}{,}
\usepackage{bm}

\usepackage[a4paper=true,driverfallback=dvipdfm,pagebackref=true]{hyperref}
\hypersetup{pdftitle = The title of my PDF, pdfauthor = My name, pdfsubject= The subject, pdfkeywords = keyword1 keyword2 keyword3} 
\hypersetup{colorlinks = true, linkcolor = green, anchorcolor = red, citecolor = blue, filecolor = red, pagecolor = red, urlcolor = red}

\begin{document}

   \title{An upgraded interpolator of the radial basis functions network for spectral calculation based on empirical stellar spectral library}
   
   \volnopage{Vol.0 (20xx) No.0, 000--000} 
   \setcounter{page}{1}

   \author{Lian-Tao Cheng\inst{1,2,3,4}, Feng-Hui Zhang\inst{1,2,3}} 

   \institute{Yunnan Observatories, Chinese Academy of Sciences, 396 Yangfangwang, Guandu District, Kunming, 650216, P. R. China; {\emph clt252206@sina.cn, zhangfh@ynao.ac.cn}\\
\and
Center for Astronomical Mega-Science, Chinese Academy of Sciences, 20A Datun Road, Chaoyang District, Beijing, 100012, P. R. China\\          
\and
Key Laboratory for the Structure and Evolution of Celestial Objects, Chinese Academy of Sciences, 396 Yangfangwang, Guandu District, Kunming, 650216, P. R. China\\      	
\and
University of Chinese Academy of Sciences, Beijing, 100049, China\\
\vs \no
   {\small Received 20XX Month Day; accepted 20XX Month Day}}

\abstract{Stellar population synthesis is an important method in the galaxy and star-cluster studies. 
In the stellar population synthesis models, stellar spectral library is necessary for the integrated spectra of the stellar population. 
Usually, the stellar spectral library is used to the transformation between the stellar atmospheric parameters and the stellar spectrum. 
The empirical stellar spectral library has irreplaceable advantages than the theoretical library. 
However, for the empirical spectral library, the distribution of stars is irregularly in the stellar atmospheric parameter space, this makes the traditional interpolator difficult to get the accurate results. 
In this work, we will provide an improved radial basis function interpolator which is used to obtain the interpolated stellar spectra based on the empirical stellar spectral library. 
For this interpolator, we use the relation between the standard variance $\sigma$ in the Gaussian radial basis function and the density distribution of stars in the stellar atmospheric parameter space to give the prior constraint on this $\sigma$. 
Moreover, we also consider the anisotropic radius basis function by the advantage of the local dispersion of stars in the stellar atmospheric parameter space. 
Furthermore, we use the empirical stellar spectral library MILES to test this interpolator. 
On the whole, the interpolator has a good performance except for the edge of the low-temperature region. 
At last, we compare this interpolator with the work in \citet{2018MNRAS.476.4071C}, the interpolation result shows an obvious improvement. 
Users can use this interpolator to get the interpolated spectra based on the stellar spectral library quickly and easily. 
\keywords{stars: fundamental parameters---stars: atmospheres---(Galaxy:) globular clusters: general---methods: numerical}
}

   \authorrunning{L.-T. Cheng et al. }            
   \titlerunning{An upgraded spectral interpolator base on RBF}  
   \maketitle

%
\section{Introduction}           
\label{sect:intro}

Stellar population synthesis model is important in astronomical research. 
Most of the observation data are from star, and these data can be used in the study of the system of stars (galaxy, the star-cluster, etc.). Stellar population synthesis model is a widely used tool in this kind of study, this is because that the stellar population is the basic ingredient in the system of stars. 
Moreover, the integrated spectrum of stellar population contains large effective information, so the calculation of integrated spectra is necessary for stellar population synthesis model. 

In the widely used evolutionary population synthesis models \citep[ etc.]{2003MNRAS.344.1000B, 2009MNRAS.398..451M, 2013MNRAS.428.3390Z}, the below three key components are used for the integrated spectra of the stellar populations. 1. initial mass function (IMF), which gives the relative number for stars with a initial mass $M$; 2. isochrone library, which is derived from the stellar evolution model and is used to give the stellar parameters (include the atmospheric parameters) for stars in a stellar population; 3. the stellar spectral library, which is used to convert the stellar atmospheric parameters to the stellar spectra.  At last, the integrated spectrum of stellar population is the sum of the stellar spectrum.

The stellar spectral library gives a correspondence between the stellar atmospheric parameters and the spectrum. Stellar spectral library can be divided to two kinds: theoretical and empirical libraries. The theoretical spectra are calculated from the stellar atmosphere model \citep[ etc.]{1992IAUS..149..225K, 2005MNRAS.357..945G}, the empirical spectra are from observations \citep[ etc.]{2001A&A...369.1048P, 2003A&A...402..433L, 2006MNRAS.371..703S, 2014A&A...565A.117C}. 

For the empirical and theoretical spectral libraries, each one has its advantages and disadvantages. Because the theoretical spectra are calculated from the atmospheric model, they have larger coverage in the wavelength, spectral resolution and the stellar atmospheric parameters range. However, they are limited by the incomplete atomic- and molecular-line lists, the uncertain abundance pattern, the assumption and idealized treatment in the model calculation, and so on \citep{2014dapb.book...63K}. Unlike the theoretical spectral library, the empirical spectra are from observations, they have the limited wavelength, resolution, noise and error that is raised from observation and the data process (e.g. flux calibration, stellar atmospheric parameter extraction). They avoid many disadvantages of the theoretical spectra. The theoretical and empirical stellar spectral libraries are complementary. 

Because the stars in the spectral library are discrete, we need an interpolator to get the spectra of any set of possible stellar atmospheric parameters. The distribution of stars in the theoretical spectral libraries have less atmospheric parameter limitations and usually is dense and regular. Under this situation, classical linear interpolation method can give a reliable interpolation result. However, for the empirical stellar spectral library, the stars are discrete in the atmospheric parameter space, this irregular distribution makes the traditional interpolator difficult to get the expected interpolation results. 
Therefor, in the work of \citet{2018MNRAS.476.4071C} we constructed an interpolator based on the radial basis function (RBF) network to get the stellar spectra in the stellar population synthesis model, the algorithm is different from those using polynomial form (e.g. \citealt{2011RAA....11..924W} and \citealt{2011A&A...531A.165P}). 

The interpolator based on the radial basis function network is called RBF interpolator in this text. The computational formula of the RBF interpolator (Eq. \ref{eq: interpolator}) is similar to the formula of field in the smooth particle hydrodynamics simulation (SPHs) and the formula of expected number in the likelihood estimation. All of them are based on the calculations of the kernel functions. In this work, Gaussian function is used as kernel function as shown in Eq. \ref{eq: kernel function}, the $\sigma$ is the standard deviation used to characterize the effect region of the kernel function.  
 
For the RBF interpolator, the kernel function does not have any strict constraint, the different settings of the kernel functions will influence the interpolation results. If all the Gaussian kernel functions have the same $\sigma$, the relatively small $\sigma$ in the sparse area will make the interpolation results be discrete, however, the relatively large $\sigma$ in the dense area will make the interpolation results be oversmoothed (lack of detailed information).
In the work of \citet{2018MNRAS.476.4071C}, the local average distance is used to give the $\sigma$ for the Gaussian kernel function. This is because the distribution of stars in the spectral library is non-homogeneous in the stellar atmospheric parameter space. 

In this work, we compare the interpolation calculation of the RBF network with the calculation of the density field in SPHs. Under this comparison, we include a constraint about the $\sigma$ in the Gaussian kernel function of RBF network from the relation between the smooth length and the density in SPHs. Under this constraint, the size of $\sigma$ for each kernel function is related to the local density of sample points in the parameter space. We use this constraint to replace the coarse determination about $\sigma$ in the work of \citet{2018MNRAS.476.4071C} for the RBF network Gaussian kernel function. 
Moreover, same as in SPHs, the spherically symmetric kernel function usually is not a better selection \citep{1983ApJ...273..749B, 1996ApJS..103..269S, 1998ApJS..116..155O}. We will refer to the process of adaptive smooth particle hydrodynamics simulation (ASHPs) and take anisotropic kernel function in the RBF interpolation calculations. 
As a result, We will present an upgraded RBF interpolator which can be used for the spectral calculation based on the empirical stellar spectral library. 

The outline of this paper is as follows. 
In Section \ref{sec: methods}, we briefly introduce the RBF network and its structure, then explain the constraint on the kernel function used in this work and the constructing anisotropic kernel function in the RBF network.
In Section \ref{sec: Optimizing}, we use the Beetle Antennae Search algorithm \citep[hereafter BAS]{DBLP:journals/corr/abs-1710-10724} to search for the best kernel function parameters. 
In Section \ref{sec: result}, we present the interpolation spectra and test this upgraded RBF interpolator by using the Medium-resolution Isaac Newton Telescope library of empirical spectra \citep[hereafter MILES,][]{2006MNRAS.371..703S, 2007MNRAS.374..664C}, and compare it with our previous work in \citet{2018MNRAS.476.4071C}. 
At last, in Section \ref{sec: conclusion}, we give the conclusion of this work.


\section{Method}
	\label{sec: methods}

In the stellar population synthesis model, the stellar spectral library provides a fast way to get the spectra of any star compared with the direct calculation by the stellar atmospheric model.
This process is a fitting or an interpolation process of the spectra in the stellar atmospheric parameter space (usually it includes three parameters: effective temperature $T_{\rm eff}$, logarithmic surface gravity acceleration ${\rm lg}\, g$ and metallicity $[Fe/H]$). The stellar spectral library gives a correspondence between stellar atmospheric parameters and stellar spectra. 

In this section, we will give a detailed introduction of the RBF network and the upgraded RBF interpolator in our work. 
In Section \ref{subsec: rbf_net}, we describe the RBF network and its calculated process as an interpolator (RBF interpolator). 
In Section \ref{subsec: interpolator}, we describe the kernel function of upgraded RBF interpolator in our work.
In Section \ref{subsec: summary of method}, we give a summary of the upgraded RBF interpolator in our work. 

\subsection{RBF network and interpolator}
	\label{subsec: rbf_net}

\begin{figure}
	\centering{
	\includegraphics[scale=0.25]{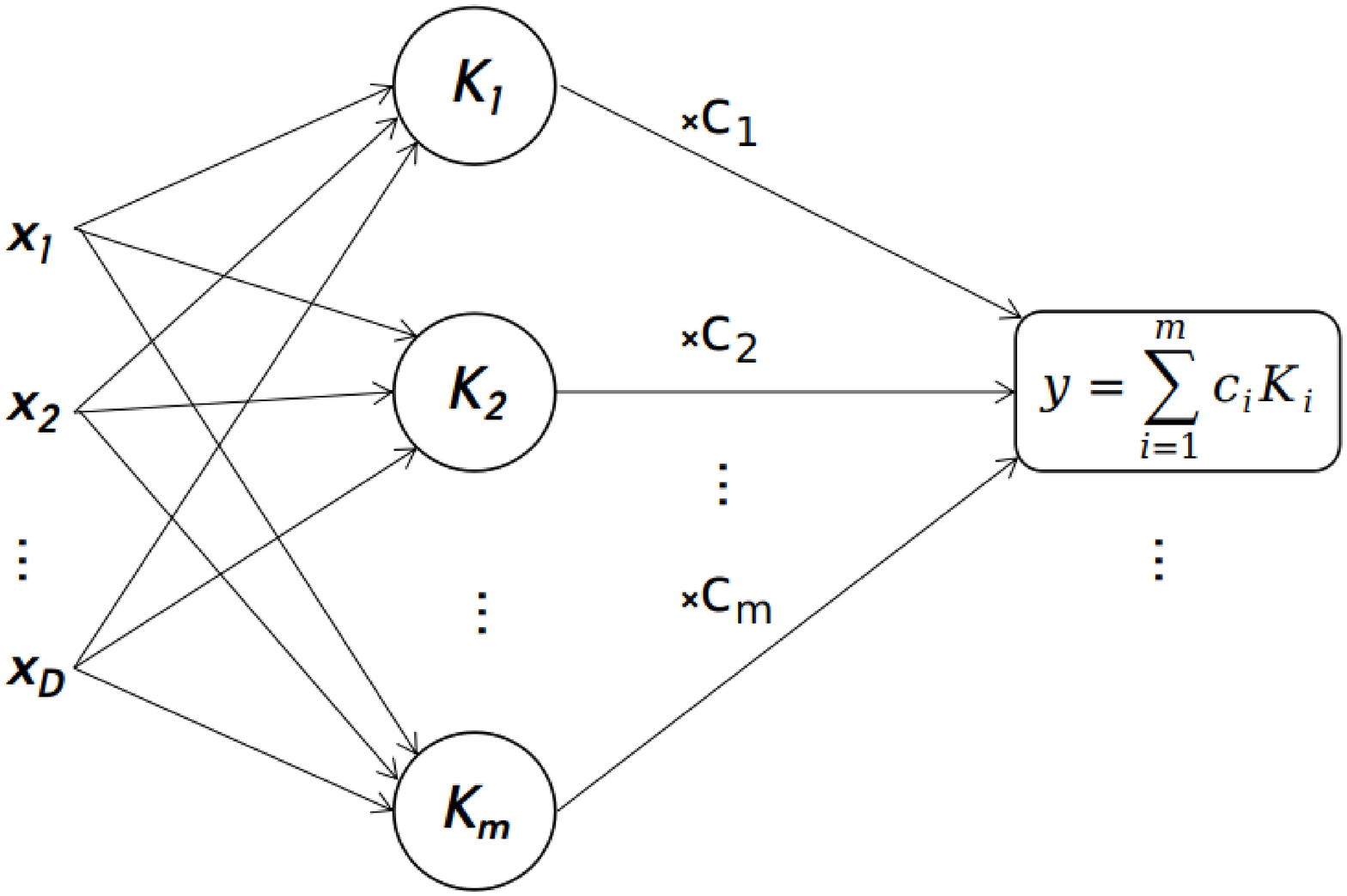}}
    \caption{The three-layer structure of RBF network is shown. The left is the input layer and $\bm{x}\ (x_1, x_2,...,x_{\rm D})$ is the input sample coordinate in the $D$-dimensional space. The middle is a hidden layer that is constituted by RBF functions $\bm{K}_i(\bm{x}-\bm{\mu}_i)$, $i = 1,2,3,...,m$, $\bm{\mu}_i\ (\mu_1, \mu_2, \mu_3,...,\mu_{\rm D})_i$ is the central coordinate of $\bm{K}_i(\bm{x}-\bm{\mu}_i)$ in the $D$-dimensional space and $m$ is the number of RBF functions in the hidden layer. The right is the output layer and $y$ is the prediction value which is the sum of RBF function multiplied by the corresponding weight factor $\bm{c}\ (c_1,c_2,c_3,...,c_m)$. 
    }
    \label{fig:net_figure}
\end{figure}

Early introduction of RBF interpolator can be found in \citet{10.5555/48424.48433}.  \cite{broomhead1988multivariable} brought the RBF into the artificial neural network (ANNs). Up to now, the RBF interpolation and fitting method has been applied widely in many fields (such as, mineral analysis, aircraft design, image processing and pattern classification). 

RBF interpolator can be thought as an application of RBF network which is a kind of kernel methods. RBF also is called kernel function in this paper. The construction of RBF network is shown in Fig. \ref{fig:net_figure}. Moreovemsxxr, the sample in this work is a set of points with  the coordinate set $\{ \bm{x}_1, \bm{x}_2,...,\bm{x}_i,...,\bm{x}_{N} \}$ and the values $\{y_1, y_2,..., y_i,..., y_{N} \}$ ($\bm{x}_i\ (x_1, x_2, ...,x_D)_i$ is the $i\,$th sample point coordinate in the $D$-dimensional space. In this paper, we use $\{ \}$ to represent set in mathematics).  

From Fig. \ref{fig:net_figure}, we can find the RBF network consists of three layers. The left is the input layer and $\bm{x}\ (x_1, x_2, ...,x_D)$ is the input point coordinate. The middle is the hidden layer which is constituted by Kernel functions $K_i(\bm{x}-\bm{\mu}_i)\ (i =1,2,...,m)$, $\bm{\mu}_i\ (\mu_1,\mu_2,\mu_3,...,\mu_{\rm D})_i$ is the central coordinate of the $i\,$th kernel function in $D$-dimensional space. The right is the output layer which is the sum of the kernel function multiplied by the weight factor ($y=\sum_{i=1}^m c_i\cdot K_i(\bm{x}-\bm{\mu_i})$) and the size of the output layer has no any limitation. In the spectral interpolation, $y$ is the flux within a given wavelength interval, it is one dimensional scalar, the interpolation spectrum consist of the interpolated fluxes at different wavelengths. 

For a sample with a huge size ($N$), a fast RBF network can be constituted by a much less number of kernel functions in the hidden layer ($m \ll N$). Usually, K-means clustering method \citep{zbMATH03340881, DBLP:conf/icml/DingH04a} is used to search for the kernel central coordinates $\bm{\mu}_i \ (i= 1,2,3,...,m)$, then linear regression method is used to get the weight factor array $\bm c\ (c_1,c_2,c_3,...,c_m)$ for the sample set $\{ (\bm{x},\ y)_i,\ i=1,2,3,...,N \}$.

The empirical stellar spectral library usually comprise several hundred or thousand spectra. So in our work, the number of kernel functions $m$ is set to be same as the sample number $N$. Moreover, we do not need the calculations of the K-means and the linear regression. We take the sample points as the centers of the kernel functions directly in the spectral RBF interpolator ($\bm{\mu}_i=\bm{x}_i,\ i=1,2,3,...,N$).
The weight factor array $\bm{c}$ of kernel function in the hidden layer is obtained by solving the system of linear equations
\begin{equation}
	\label{eq: linear equations}
	\sum_{j=1}^N K_j(\bm{x}_i-\bm{x}_j)\cdot c_j = y_i \ (i = 1, 2, 3,..., N),
\end{equation}
In our works the Gaussian kernel function ${\rm e}^{-\sum_{d=1}^D(x_d - \mu_d)^2/(2 \sigma^2)}$ is used,
\begin{equation}
	\label{guass}
	K_j(\bm{x}_i-\bm{x}_j) = {\rm e}^{\frac{-\sum_{d=1}^D(x_{d,i}-x_{d,j})^2}{2\sigma_j^2}}, 
\end{equation}
where $\bm{x}_{d,i}$ and $\bm{x}_{d,j}$ are the coordinates of the $i\,$th and $j\,$th sample point in the $D$-dimensional space ($d=1,2,3,...,D$). The central coordinate of the $j\,$th kernel function $K_j$ is $\bm{x}_j$, $\sigma_j$ is the standard deviation of the $j\,$th kernel function $K_j$ and can be used to characterize the influence range of the kernel function. $c_j$ is the weight factor of the $j\,$th Gaussian kernel function. 
Solving the linear Eq. \ref{eq: linear equations}, we can get the weight factor array $\bm{c}$, if sample coordinate in set $\{ \bm{x}_1, \bm{x}_2, \bm{x}_3,...,\bm{x}_{N} \}$ is different from each other \footnote{For Gaussian kernel function, this conclusion can get a fixed factor array $\bm{c}$ from the Micchelli theorem \citep{Micchelli1986}}. After obtaining $\bm c$, we can get a simple formula for the interpolation calculation,
\begin{equation}
	\label{eq: interpolator}
	y(\bm{x})=\sum_{j=1}^N K_j(\bm{x}-\bm{x}_j)\cdot c_j,
\end{equation}
where $\bm{x}$ is the input coordinate, and $y(\bm{x})$ is the interpolation result. For spectrum interpolation calculation, $\bm{x}$ and $y$ correspond to the stellar atmospheric parameter and the interpolated flux within a wavelength interval, respectively.

The RBF network is base on the kernel functions. However, a const $\sigma$ usually is used for all the Gaussian kernel functions \citep{51953}. For the irregular distribution of stars in the stellar spectral library, all the kernel functions with the same $\sigma$ is not a better selection. Next, we will introduce the kernel function in our spectral RBF interpolator.

\begin{figure}
	\centering{
	\includegraphics[scale=0.5]{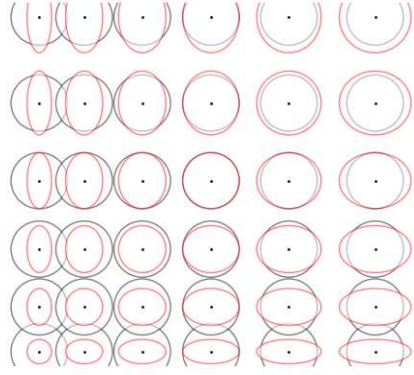}}
    \caption{A example of the 2-D space $\sigma$ distribution for the kernel function. Black points are the centers of the kernel functions ($\{ \bm{\mu}_i, i=1,2,3,...,m \}$). From the left to the right and from the bottom to the top, the density of points decrease. The radius of black rings is used to characterize the $\sigma$ influence range for the classical RBF network. The red ellipse is used to characterize a more reasonable influence range of kernel functions for anisotropic distribution.}
    \label{fig: example}
\end{figure}

\subsection{Gaussian kernel function in the upgraded RBF interpolator}
	\label{subsec: interpolator}
 
The const $\sigma$ in the Gaussian kernel functions has a disadvantage: large $\sigma$ of the kernel function makes the interpolation results oversmoothed in the dense area, the relative small $\sigma$ makes the interpolation results discrete in the sparse area. 
In Fig. \ref{fig: example}, we give a example of the 2-D $\sigma$ distribution, black point is the centers of the kernel functions $\{ \bm{\mu}_i, i=1,2,3,...\}$ , and the space density decreases from the left to the right and the bottom to the top, and the anisotropic distribution exists in top-left and bottom-right parts. The black rings are used to characterize the influence range of kernel functions in the traditional RBF network and the red  ellipses show a better choice for the anisotropic kernel functions. 

Stars in the empirical spectral library face a more complex situation than Fig. \ref{fig: example}. The distribution is nonuniform and its density varies significantly in the stellar atmospheric parameter space (a typical situation can be found in Fig. \ref{fig: isotropy_sigma}). This situation is resulted from the observational and theoretical limitations \footnote{For spectroscopic observations, only solar neighbor stars can be observed with high quality. In stellar theory, the bright stars allows have a short evolutionary time scale. Both of them make the observed spectral sample have obvious selection effect}.   
In our work, we set a constraint by including a relation between the $\sigma$ and the spatial density of sample points into the RBF network, this relation can be used to determine the $\sigma$ value. Moreover, we also consider the anisotropy of the kernel function in the RBF network by referring to the adaptive smoothed particle hydrodynamics simulation (hereafter ASPHs). 

In Section \ref{subsec: constraint}, we give an introduction of the constraint on $\sigma$. In Section \ref{subsec: anisotropy}, we show generic Gaussian kernel function for the anisotropic situation in our works.

\subsubsection{The constraint on $\sigma$ and its size calculation} 
	\label{subsec: constraint}	

In this part, we introduce the smooth length constraint of SPHs into the RBF network. Under this constraint, we show the computing method of $\sigma$ for the kernel function in the RBF network.  

In the SPHs, the sample consists of particles with the coordinate set $\{ \bm{x}_i,\ i = 1,2,3,...,N \}$ and mass set $\{ m_i,\ i = 1,2,3,...,N \}$, $\bm{x}_i$ and $m_i$ are the position and mass of the $i\,$th particle. The fluid density $\rho$ in the position $\bm{x}_i$ is
\begin{equation}
	\label{eq: density_sph}
	\rho_i \approx \sum_{j=1}^{N} m_j \cdot W_{ij} \ (i =1,2,3,...,N),
\end{equation}
where $N$ is the number of particles, $W_{ij}$ is $a\cdot K(\bm{x}_i-\bm{x}_j)$ and $a$ is the normalized coefficient. We use the Gaussian kernel function
\begin{equation}
	\label{eq: kernel function}
	W_{ij} = \frac{1}{(\sqrt{2\pi}\sigma_j)^D} \cdot {\rm e}^{\frac{-\sum_{d=1}^D(x_{d,i}-x_{d, j})^2}{2\cdot \sigma_j^2}}, 
\end{equation}
where $D$ is the space dimension and $\sigma_j$ is the smooth length of the $j\,$th kernel function.
Replacing $m_j$ with $\rho_jV_j$, where $V_j$ is the volume of $j\,$th particle in SPHs. We have
 \begin{equation}
	\label{eq: density_likehood}
	\rho_i \approx \sum_{j=1}^{N} \rho_j\cdot V_j \cdot W_{ij} = \sum_{j=1}^{N} \rho_j\cdot V_j \cdot \frac{1}{(\sqrt{2\pi}\sigma_j)^D} \cdot {\rm e}^{\frac{-\sum_{d=1}^D(x_{d,i}-x_{d, j})^2}{2\cdot \sigma_j^2}}.
\end{equation} 

The fluid density varies with time. A better smooth length should vary dynamicly. Usually $\sigma = \sigma_0 \cdot (\rho_0 / \rho)^{1/D}$ is used to give the current smooth length ($\sigma_0$ and $\rho_0$ are initial value of SPHs). This relation gives a constraint about the $\sigma$ by spatial density of sample points. Here, we use this relation in the kernel function of the RBF network at different positions (in the stellar atmospheric parameter space). We give the constraint about $\sigma$,
\begin{equation}
	\label{eq: dens-sigma}
	\sigma  \propto \rho^{\frac{-1}{D}} \propto V^{\frac{1}{D}},
\end{equation}
where $V$ is used to characterize the particle volume which is related to the influence range of the kernel function. Usually, the dimension number $D$ is 3 for the stellar spectral library. 

Now, including the constraint (formula \ref{eq: dens-sigma}) in Eq. \ref{eq: density_likehood}, we have
\begin{equation}
	\label{eq: dens}
	\rho_i \approx \sum_{j=1}^N \rho_j \cdot \frac{c_0}{(\sqrt{2\pi})^D} \cdot {\rm e}^{ \frac{-\sum_{d=1}^D(x_{d,i}-x_{d, j})^2}{2\cdot \sigma_j^2}} \ (i = 1,2,3,...,N),
\end{equation} 
where $c_0 \equiv V/(\sigma^D)$ .  Eq. \ref{eq: dens} gives a set of nonlinear equations of $\sigma $. However, it is difficult to be solved. 

To simplify the Eq. \ref{eq: dens}, we approximate Eq. \ref{eq: dens} to the following expression by replacing the $j$ with $i$ for the subscripts of $\sigma$ and $\rho$,
\begin{equation}
	\label{eq: dens-dens}
	1 \approx \sum_{j=1}^N \frac{c_0}{(\sqrt{2\pi})^D} \cdot {\rm e}^{\frac{-\sum_{d=1}^D(x_{d,i}-x_{d, j})^2}{2\cdot \sigma_i^2}} \ (i = 1,2,3,...,N),
\end{equation}  
this is because $\rho$ and $\sigma$ are continuous and the adapted kernel function (here, it is Gaussian function) is for local region. 
The equation becomes independent of each other. 
From Eq. \ref{eq: dens-dens}, we can use the bisection method to calculate $\sigma_i (i=1,2,3,...,N)$ quickly. 
In this process the $\sigma$ can be adjusted by the control parameter $c_0$. From Eq. \ref{eq: dens-dens}, we can know $c_0$ is within the interval of $( (2\pi)^{D/2}/{N},(2\pi)^{D/2}\rbrack$, its upper and lower limits correspond to $\infty$ and $0$ for $\sigma$.

\begin{figure}
	\centering{
	\includegraphics[scale=0.5]{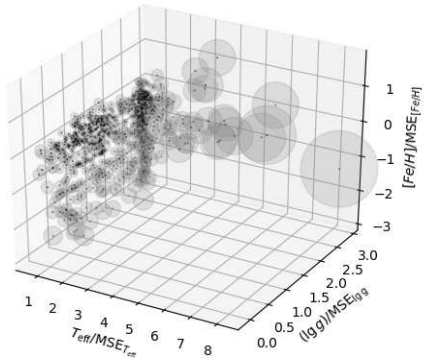}}
    \caption{The size of $\sigma$ in the kernel function for the MILES stellar spectral library. The coordinates of black points are the dimensionless stellar atmospheric parameters of MILES library ($T_{\rm eff}/{\rm MSE}_{T_{\rm eff}}$, $({\rm lg}\, g) /{\rm MSE}_{{\rm lg}\,g}$ and $[Fe/H]/{\rm MSE}_{[Fe/H]}$, the $\rm MSE$ is the mean square error function). The radius of semitransparent sphere around the black point characterizes the corresponding $\sigma$.}
   \label{fig: isotropy_sigma}
\end{figure}

In Fig. \ref{fig: isotropy_sigma}, we show the resulted $\sigma_i$ \ ($i=1,2,3,...,N$) for the dimensionless stellar atmospheric parameter of MILES library. In this work, the coordinates $(T_{\rm eff}, {\rm lg}\, g, [Fe/H])$ are dimensionless which are expressed in the units of mean square error (hereafter $\rm MSE$). The size of $\sigma$ is relatively large in the spares area and small in the dense area. 

We should also notice that the $\sigma_i$ in Eq. \ref{eq: dens-dens} is scalar, which corresponds to the isotropic kernel function. In fact, the anisotropic distribution of stars in the stellar atmospheric parameter space is common for empirical spectral library (shown in Fig. \ref{fig: isotropy_sigma}). 
 
\subsubsection{Anisotropic Gaussian kernel function}
	\label{subsec: anisotropy}
This part gives the introduction of the anisotropic kernel function by generalizing the Gaussian kernel function.  

In SPHs, the fluid varies with time and the deformation usually is not isotropy. For the anisotropy deformation, the isotropy kernel function usually is not able to give a better smoothing effect as shown in Fig. \ref{fig: example}.
Similarly, \citet{1983ApJ...273..749B} has even considered the kernel function with different smoothing length in y and z axis direction in the tidal destroy simulation. Moreover, the generalized anisotropy kernel function has been discussed by \citet{1996ApJS..103..269S} and \citet{1998ApJS..116..155O}. 

The anisotropic kernel function is the generalization of the isotropic Gaussian kernel function. If we let $\bm{M}_{\rm iso}^2$ be $\sum_{d=1}^3 (x_d - \mu_d)^2/(2 \sigma^2)$, which is the exponential part of the isotropic Gaussian kernel function, then the corresponding vector $\bm{M}_{\rm aniso}$ of anisotropic kernel function can be described by the following
\begin{eqnarray}
	\label{eq: anisotropy matrix}
	\bm{M}_{\rm aniso} & =& \bm{T} \cdot (\bm{x}-\bm{\mu}) = \sum_{k=1}^3 T_{dk} \cdot (x_k - \mu_k) \\
	 	& =& {\left[ \begin{array}{ccc}
	T_{11} & T_{12} & T_{13}\\ 
	T_{21} & T_{22} & T_{23}\\ 
	T_{31} & T_{32} & T_{33}
	\end{array} 
	\right ]}  
	{\left[ \begin{array}{c}
	x_1 - \mu_1 \\
	x_2 - \mu_2 \\
	x_3 - \mu_3
	\end{array}
	\right ]} \notag \\
	& =& {\left[ \begin{array}{ccc}
	Z_{11} & 0 & 0\\ 
	0 & Z_{22} & 0\\ 
	0 & 0 & Z_{33} 
	\end{array} 
	\right ]} 
	{\left[ \begin{array}{ccc}
	R_{11} & R_{12} & R_{13}\\ 
	R_{21} & R_{22} & R_{23}\\ 
	R_{31} & R_{32} & R_{33} 
	\end{array} 
	\right ]}
	{\left[ \begin{array}{c}
	x_1 - \mu_1 \\
	x_2 - \mu_2 \\
	x_3 - \mu_3
	\end{array} 
	\right ]} \notag \\
	(d & =& 1,2,3), \notag
\end{eqnarray}
where $d$ and $k$ is the dimensional subscript, the matrix $\bm{T}$ is a linear translation (from $\bm{x}-\bm{\mu}$ to $\bm{M}_{\rm aniso}$). The generalized Gaussian kernel function result can be written as ${\rm e}^{-\bm{M}_{\rm aniso}^2}$. 

Matrix $\bm{T}$ can be separated into two parts $\bm{T}=\bm{Z}\cdot \bm{R}$ as shown in Eq. \ref{eq: anisotropy matrix}.
The diagonal matrix $\bm{Z}$ is used to give a scale transform along with axis, the matrix $\bm{R}$ is constituted by three orthogonal basis and gives a rotational transform. Both $\bm{Z}$ and $\bm{R}$ change with the position of kernel function center $\bm{\mu}$ in the stellar atmospheric parameter space.
The isotropy kernel function can be recovered by setting $\bm{I}=\bm{R}=\sqrt{2\sigma}\cdot \bm{Z}$, here the $\bm{I}$ is the unit matrix.

In the stellar spectral library, most of stars distribute along the main sequence and in the red giant region, these two parts are distributed almost along the $T_{\rm eff}$ and ${\rm lg}\,g$ axis direction, the distribution of stars at different $[Fe/H]$ range only have a slight bias. Therefore, in most cases, the anisotropy is in the axis direction. To simplify the calculation process, in this work we ignore the rotational matrix $\bm{R}$ by setting $\bm{R} = \bm{I}$ and only consider the $Z_{\rm 11},\ Z_{\rm 22}$ and $Z_{\rm 33}$ in the generic Gaussian kernel function.

The degree of anisotropy of kernel function is dependent of the ratio of $Z_{\rm 11},\ Z_{\rm 22}$ and $Z_{\rm 33}$.  In our work, the axial local dispersion $\bm D$ is used to give the $\widetilde{\bm Z}$ for any kernel function with central coordinate $\bm \mu$,
\begin{equation}
	\label{eq: local dispersion}
	\begin{aligned}
	Z_{dd} \propto \widetilde{Z}_{dd} &= c_1+ \frac{(1-c_1)D_{d}}{(\sum_{d=1}^3 D_{d}^2)^{1/2}} \ (d=1,2,3), \\
	D_{d} &= \sqrt{\sum_{i=1}^N (x_{d,i}-\mu_d)^2 \cdot {\rm e}^{\frac{-\sum_{d=1}^3 (x_{d,i}-\mu_{d})^2}{2 \sigma^2}}}
	\end{aligned}
\end{equation}
where $c_1$ is a control parameter which is used to adjust the degree of anisotropy, the Gaussian function is used to limit the calculation in the local region. So after the calculation of Eq. \ref{eq: local dispersion}, we can know the ratio of $Z_{\rm 11},\ Z_{\rm 22}$ and $Z_{\rm 33}$ in the matrix $\bm{Z}$ for all kernel functions and get the corresponding normalized matrix $\widetilde{\bm{Z}}$. The last $\bm{Z}$ is given by solving the Eq. \ref{eq: dens-dens} again with the anisotropic kernel function ${\rm e}^{-[\bm{\widetilde Z}(\bm{x-\mu})]^{2}}$.

\subsection{Summary of the method}
	\label{subsec: summary of method}
Here, we give a summary of the mathematical process of the RBF interpolator in this work. 
In general, as shown in Fig. \ref{fig:net_figure}, there are two key parts for the RBF network need to be given, one is the kernel functions $\{ K_1, K_2, K_3,...,K_N \}$, another is the coefficients array $\bm{c}$. For the latter, $\bm{c}$ can be obtained by solving the system of linear equations \ref{eq: linear equations} for the given kernel functions and the sample $\{ (\bm{x}, y)_i, i=1,2,3,...,N \}$. Next, we will list the determination process of the RBF kernel functions $\{ K_1, K_2, K_3,...,K_N \}$.

First of all, we use the stellar spectral library as the sample of RBF network, $\{ {\bm x}_1,  {\bm x}_2, {\bm x}_3,..., {\bm x}_N,\}$ corresponds to the stellar atmospheric parameter and $\{y_1, y_2, y_3,...,y_N \}$ corresponds to the flux. We select the Gaussian kernel function and set the central coordinates of all kernel functions are the coordinates of the sample ($\bm{\mu}_i=\bm{x}_i, i=1,2,3,...,N $). Next, for the $i\,$th kernel function ($i=1,2,3,...,N$), three steps to be executed.
\begin{enumerate}
\item Solving equations \ref{eq: dens-dens} to get $\sigma_{{\rm iso},i}$ of the isotropic Gaussian kernel function. In this step, we include parameter $c_0$ for all points. 

\item Inputting $\sigma_{{\rm iso},i}$ in Eq. \ref{eq: local dispersion} and calculating the matrix $\widetilde{\bm{Z}}_i$. In this step, $c_1$ is used to control the degree of anisotropy of kernel functions. 

\item Rewriting the exponential part of the equations \ref{eq: dens-dens} by ${\rm e}^{-[\bm{\widetilde Z}(\bm{x-\mu})]^{2}}$ and solving it, we can get the last $\bm{Z}_i$ and $\bm{T}_i$ (in this step, a new parameter $c_2$ is included which corresponds to $c_0$ in the first step).

\end{enumerate}
After those three steps, we get the last kernel function of RBF network, then solve the system of equations \ref{eq: linear equations} to get the weight factor array $\bm c$. At last, we get the last RBF network, and the calculation of spectral RBF interpolator corresponds to Eq. \ref{eq: interpolator}.

The three control parameters $c_0, c_1,c_2$ are included, the optimization calculation of these parameters is shown below. For distinguishing the RBF interpolator in \citet{2018MNRAS.476.4071C} from that in this work, we name there two spectral RBF interpolators $\rm RBF18$ and $\rm RBF_{update}$.

\begin{figure*}
	\centering{
	\includegraphics[width=1\textwidth]{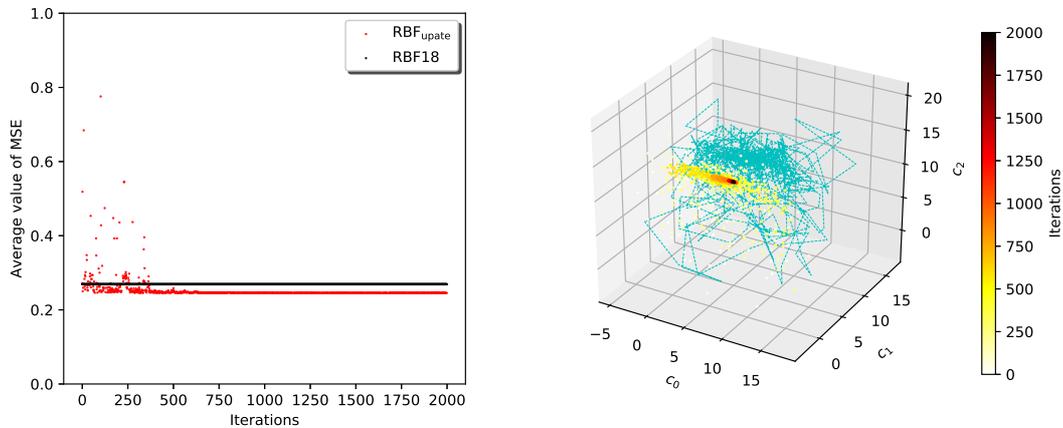}}
    \caption{The output of the optimization process by the BAS search algorithm. The left panel shows the variation of object function as a function of the iteration times. Red point is the objective function value of the $\rm RBF_{update}$ interpolator. Black point is the objective function value of $\rm RBF18$ interpolator which is used as a comparison. The right panel shows the corresponding track of the point $(c_0,\ c_1,\ c_2)$. The cyan dash line is displacement, and nodes are the coordinates of $(c_0,\ c_1,\ c_2)$ in different iterations. The node color corresponds to the number of iterations, 2000 iterations are shown here.}
   \label{fig: iterations}
\end{figure*}

\section{Optimizing of the RBF network control parameters}
	\label{sec: Optimizing}

\begin{figure*}
	\centering{
	\includegraphics[width=1\textwidth]{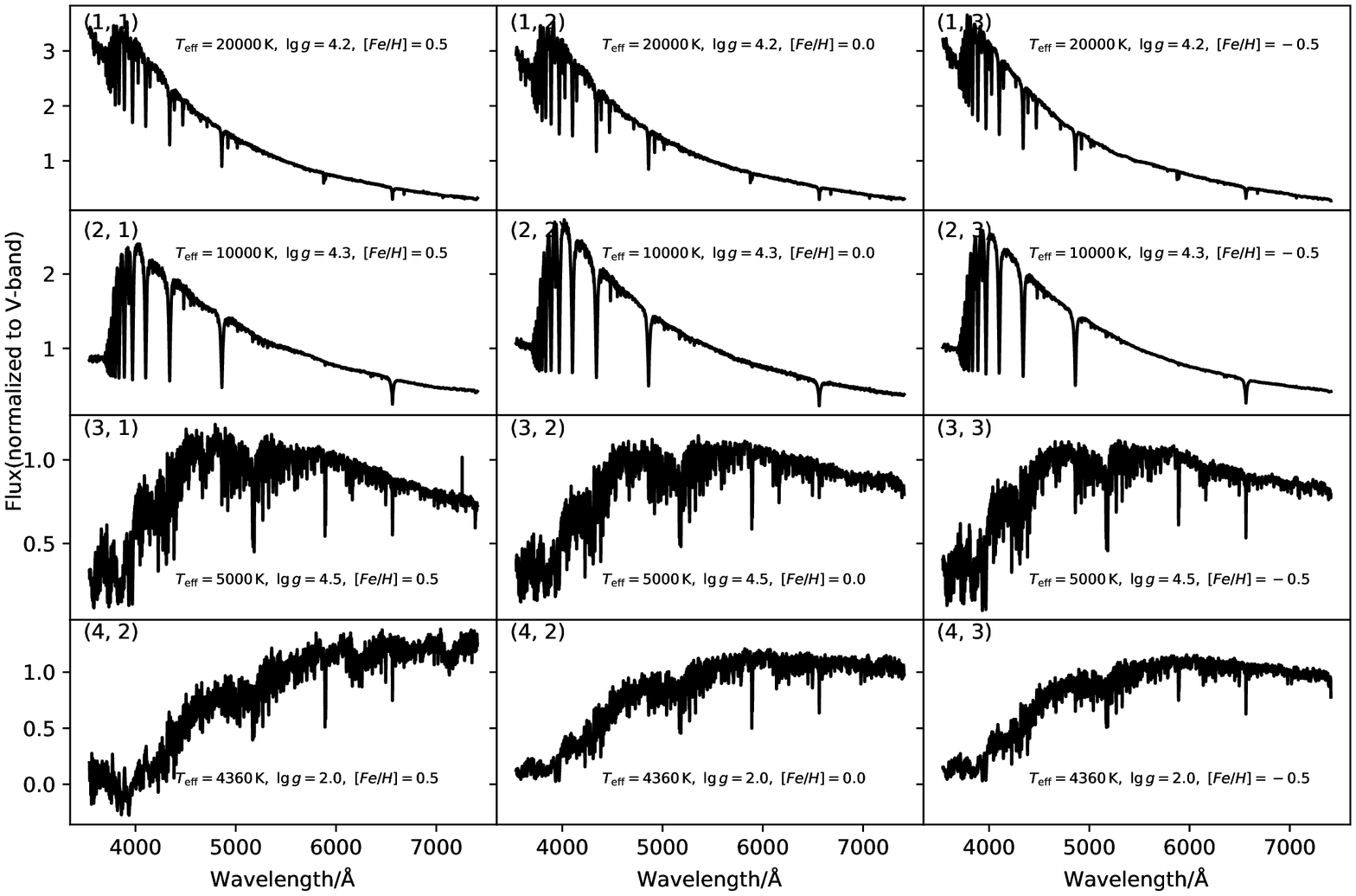}}
    \caption{The twelve interpolation spectra based on the MILES spectral library by the $\rm RBF_{update}$ interpolator. Input stellar parameters are shown in each panel. Here the spectra are dimensionless. The first two lines show the interpolation spectra of six main sequence stars with $T_{\rm eff}=20000\,\rm K $ and $T_{\rm eff}=10000\,\rm K$, The third line shows the interpolation spectra of three main sequence stars with $T_{\rm eff}=5000\,\rm K$. The forth line shows the interpolation spectra of three red giant stars with $T_{\rm eff}=4360\,\rm K$. In each line, the input stellar metallicity is $\rm [Fe/H] = 0.5, 0.0 $ and $-0.5$.}
   \label{fig: results}
\end{figure*}

We can control the RBF network by adjusting the control parameters $(c_0,\ c_1,\ c_2)$ (Section \ref{subsec: summary of method}). 
In this section, we will introduce the optimization process of these three control parameters.

In this work, MILES library is used to build the spectral RBF interpolator $\rm RBF_{update}$. The MILES empirical stellar spectral library includes $\sim 1000$ stars obtained on the $2.5\,\rm m$ Isaac Newton Telescope. The wavelength ranges from $3540.5$ to $7409.6\,{\rm \AA }$ and the spectral resolution is $\sim 2.3\, {\rm \AA }$ \citep[FWHM]{2006MNRAS.371..703S}. The coverage of the stellar atmospheric parameters are: $2748 < T_{\rm eff} < 36 000\, {\rm K}$, $ 0.00 < {\rm lg}\,g \rm < 5.50$ and $-2.93 < [Fe/H] < +1.65$. The MILES spectral library has a larger coverage in the parameter spaces than the other empirical stellar spectral libraries used in the stellar population synthesis models \citep{2007MNRAS.374..664C}.

In this work, we use the semi-empirical BaSeL-3.1 stellar spectral library \citep{1997A&AS..125..229L, 1998A&AS..130...65L, 2002ASPC..274..166W} as the reference library to find the best control parameters of $\rm RBF_{update}$ interpolator. BaSeL-3.1 is one of the widely used spectral libraries, it provides an extensive and homogeneous grid of low-resolution spectra in the range of $91-1600 000\,\rm \AA$ for a large range of stellar parameters: $2000 < T_{\rm eff} < 50 000\,\rm K$; $-1.02 < {\rm lg}\, g < 5.5$ and $-5.0 < [Fe/H] < 1.0$.

For avoiding extrapolation of the spectral RBF interpolator, the input parameter should be within the coverage area of stars in the MILES library. So in the BaSeL-3.1 library only those models within the coverage area of MILES library are used here. Gaussian smooth algorithm is used to degrade the resolution of the output spectra to $20\, \rm \AA$ (the resolution of BaSeL-3.1 in visible wavelength). Iterative calculation is used in the optimization, in each iteration, hundreds of stars in BaSeL-3.1 are selected as the input sample, the output interpolation spectra are used to compare with the original spectra in BaSeL-3.1. The average value of the mean square error between interpolation spectra and original spectra in BaSeL-3.1 are used as objective function $f(c_0, c_1,c_2)$ in the optimizing process, and the best $(c_0, c_1,c_2)$ corresponds to the minimum of $f(c_0, c_1,c_2)$. 

The widely used optimization algorithm needs massive computation to obtain the best  $(c_0, c_1,c_2)$. BAS algorithm is used \citep{DBLP:journals/corr/abs-1710-10724}, which is a new and light algorithm by simulating the beetle behavior. 
In this work, the process of the BAS search is in 3-D space of $(c_0, c_1,c_2)$ and comprises four steps.
\begin{description}
	\item{Step1.} Setting the initial position of the 'beetle' $\bm{P}_0$, the initial distance of two antennas $A_0$ and the initial step length $S_0$. For the $i\,$th iteration, it is referred to as $\bm{P}_i$, $A_i$, $S_i$;
	\item{Step2.}  Generating a unit vector $\bm{d}$ with random direction, $\bm{d}$ is used to give the relative position of two antennas. Then, left antenna position has $\bm{P}_{\rm l, i}=\bm{P}_i - 0.5\cdot A_i\cdot \bm{d}$ , right antenna position has $\bm{P}_{{\rm r},i}=\bm{P}_i+ 0.5\cdot A_i\cdot \bm{d}$;
	\item{Step3.} Calculating the objective function on two antennas $f(\bm{P}_{{\rm l},i}), f(\bm{P}_{{\rm r},i})$. The new position is $\bm{P}_{\rm new} = \bm{P}_i+[f(\bm{P}_{{\rm r}, i})- f(\bm{P}_{{\rm l},i})]/ {\rm abs}(f(\bm{P}_{{\rm r},i})- f(\bm{P}_{{\rm l},i})) \cdot \bm{d} \cdot S_i$;
	\item{Step4.} If iteration meets the critical condition, the process jump out of the iterations. If not, $i\,+=\,1$, $\bm{P}_i = \bm{P}_{\rm new}$ and the loop go back to Step2.  
\end{description}
During iteration, the distance of two antennas $A_i$ and the step length $S_i$ change slowly. In our calculation, $S_{i+1}=b \cdot S_i$ and $A_i = c\cdot S_i$, $b$ is a constant coefficient  close to 1 but less than 1 and $c$ also is a constant coefficient. 

Fig. \ref{fig: iterations} gives the iteration process of the $(c_0, c_1, c_2)$ in the optimization. In this figure, 2000 iterations are shown. The left panel shows the objective function output, red points are the objective function values of $\rm RBF_{update}$ interpolator. As a comparison, the black points are the objective function value of $\rm RBF18$ interpolator with the same calculation. In the right panel, the cyan dash line shows the movement trail of 'beetle', and the color nodes are the 'beetle' positions in iteration. The node color corresponds to the number of iterations. 
We can find the result of $(c_0, c_1,c_2 )$ converges to a fixed value at last. The objective function of $\rm RBF_{update}$ interpolator is lower than the result of $\rm RBF18$ interpolator, this means that spectral interpolator $\rm RBF_{update}$ has a better performance than $\rm RBF18$ in spectral interpolation calculation of the stars in BaSeL-3.1 library.

\begin{figure}
	\centering{
	\includegraphics[scale=0.5]{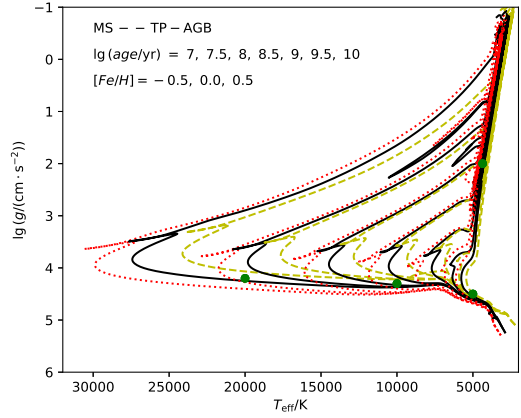}}
    \caption{The positions (green points) of the stars in Fig. \ref{fig: results}. The isochrones are used to give the relative position. For each isochrone, stars from the zero-age main sequence to the end of TP-AGB phases are displayed. The red dot lines have $\rm [\emph{Fe/H}]=-0.5$, black lines have $\rm [\emph{Fe/H}]=0$ and yellow dot dash lines have $\rm [\emph{Fe/H}]=0.5$. From left to right $\rm lg\, (\emph{age}/yr)$ are $7,\ 7.5,\ 8,\ 8.5,\ 9,\ 9.5$ and 10.}
   \label{fig: isochrone}
\end{figure}

\section{Result and analysis}
	\label{sec: result}
	In this section, we show the interpolation results and the analysis of $\rm RBF_{update}$ interpolator based on the empirical stellar spectral library MILES.
In Section \ref{subsec: results}, we present the interpolated stellar spectra for different spectral types. 
In Section \ref{subsec: test_result}, we give a test of the $\rm RBF_{update}$ interpolator and give a comparison between the interpolation spectra by $\rm RBF_{update}$ and $\rm RBF18$ interpolators based on the MILES stellar spectral library.
In Section \ref{subsec: analysis}, we give an analysis and a discussion for the behave of the $\rm RBF_{update}$ interpolator in our test. 	

\subsection{Result}
	\label{subsec: results}
For the sake of clarity, we only show the interpolation spectra of twelve stars with typical stellar atmospheric parameters in Fig. \ref{fig: results}. All of the interpolation spectra are normalized to its V-band flux. The twelve panels are divided in to four rows, and each row corresponds to three input stars with different metallicity $[Fe/H] = -0.5,\ 0$ and $0.5$ (the corresponding atmospheric parameters is shown in each panel). In the first row, the input stars with $T_{\rm eff}=20000\, \rm K$ and $\rm lg\,\emph{g}=4.2$ correspond to the massive main-sequence star and have a great effect on the U-band flux of the stellar population integrated spectra with stellar population age less than $\rm 10^{7.5}\, \rm yr$. In the second row, the input stars with $T_{\rm eff} = 10000\,\rm K$ and $\rm lg \, \emph{g}=4.3$ correspond to medium mass main-sequence star and have a great effect on the B-band flux of integrated spectra with stellar population age less than $\rm 10^{8.5}\,\rm yr$. In the third row, the input stars with $T_{\rm eff} = 5000\,\rm K$ and $\rm lg\, \emph{g} = 4.5$ correspond to the low-mass main-sequence star which have a large number in the stellar population, they have a significant effect on the V-band of the stellar populations integrated spectra. In the last row, the input stars with $T_{\rm eff} = 4360\,\rm K$ and $\rm lg\,\emph{g} = 2.0$ correspond to red giant stars. It is very bright and have a great effect on the infrared band flux of the older stellar population integrated spectra. 

In Fig. \ref{fig: isochrone}, we give the positions of these stars (in Fig. \ref{fig: results}) on the $T_{\rm eff}$ and ${\rm lg}\,g$ plane (the green points). Moreover, we present three sets of the isochrones at metallicity $[Fe/H] = -0.5,\ 0$ and $0.5$ to give the locations of the stellar populations. For each set of isochrones, from left to right the stellar age ${\rm lg\,(age}/{\rm yr)}=7,\ 7.5,\ 8,\ 8.5,\ 9,\ 9.5$ and $10$, The isochrones only show the stars from the zero-age main sequence phase to the end of thermal pulsating asymptotic giant branching (TP-AGB) phase. The relative positions between stars and the isochrones correspond to the above analysis of Fig. \ref{fig: results}.
In this figure, the isochrones are the results of MIST \citep{2016ApJS..222....8D, 2016ApJ...823..102C}, they are calculated by using the stellar evolution code MESA \citep{2011ApJS..192....3P, 2013ApJS..208....4P, 2015ApJS..220...15P, 2018ApJS..234...34P}.

\begin{figure*}
	\centering
	\includegraphics[width=1\textwidth]{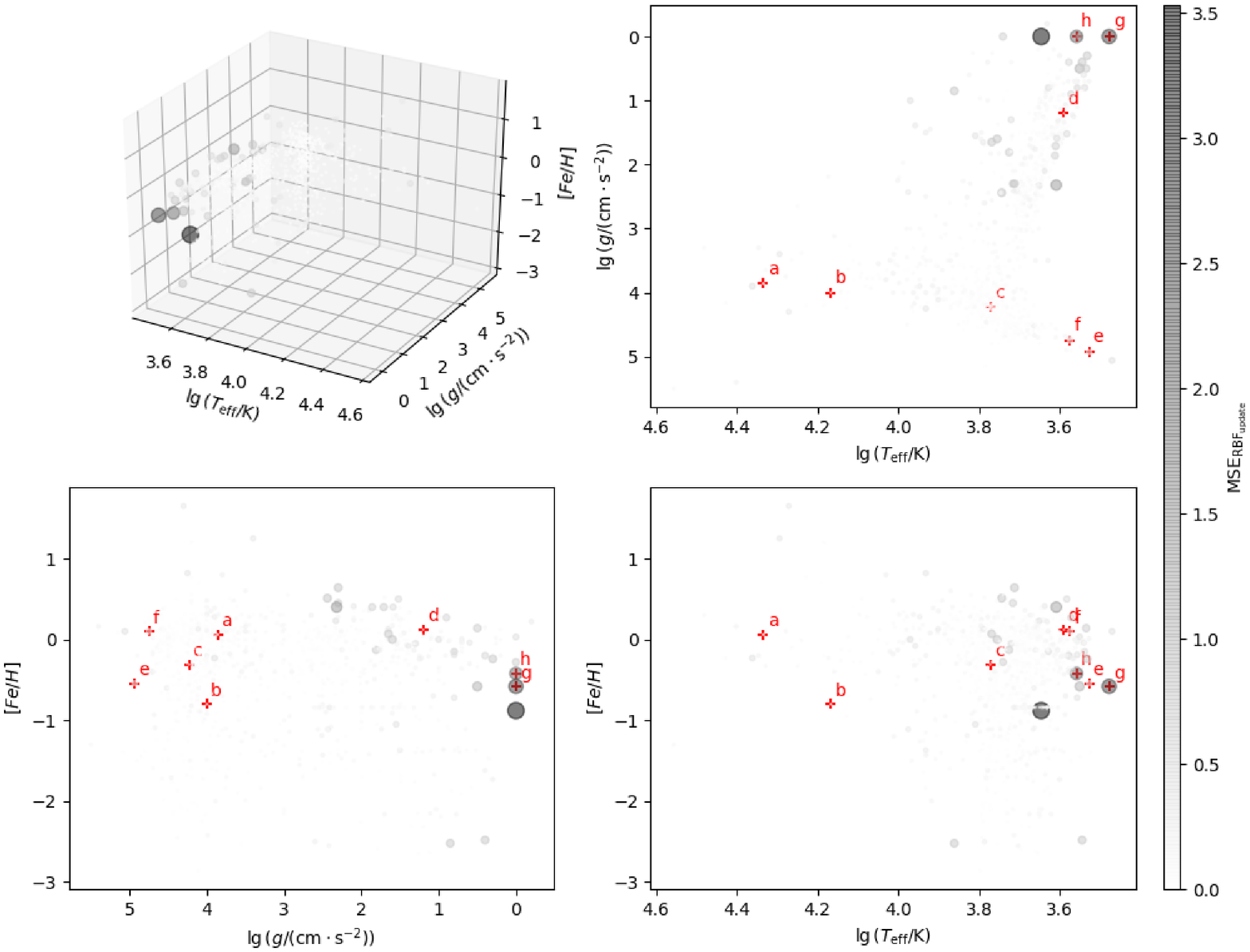}
    \caption{The $\rm MSE_{RBF_{update}}$ based on MILES spectral library. The top-left panel gives the result in stellar atmospheric parameter space. The points give the positions of test stars, the grey level and point size are used to characterize the test result $\rm MSE$ of $\rm RBF_{update}$ interpolator. Here, $\rm MSE$ is the mean square error between the interpolated and the original spectra of test star. The top-right panel gives the projection on the ${\rm lg}\,T_{\rm eff}$ and ${\rm lg}\,g$ plane, the bottom-left panel gives the projection on the ${\rm lg}\,g$ and $[Fe/H]$ plane and the bottom right panel gives the projection on the ${\rm lg}\,T_{\rm eff}$ and $[Fe/H]$ plane. The red "+" and the corresponding letter are used to give the position of the stars for which their spectra are shown in Fig. \ref{fig: spec_comp}.
    }
   \label{fig: MSE_distribution}
\end{figure*}	
		
\begin{figure*}
	\centering
	\includegraphics[width=1\textwidth]{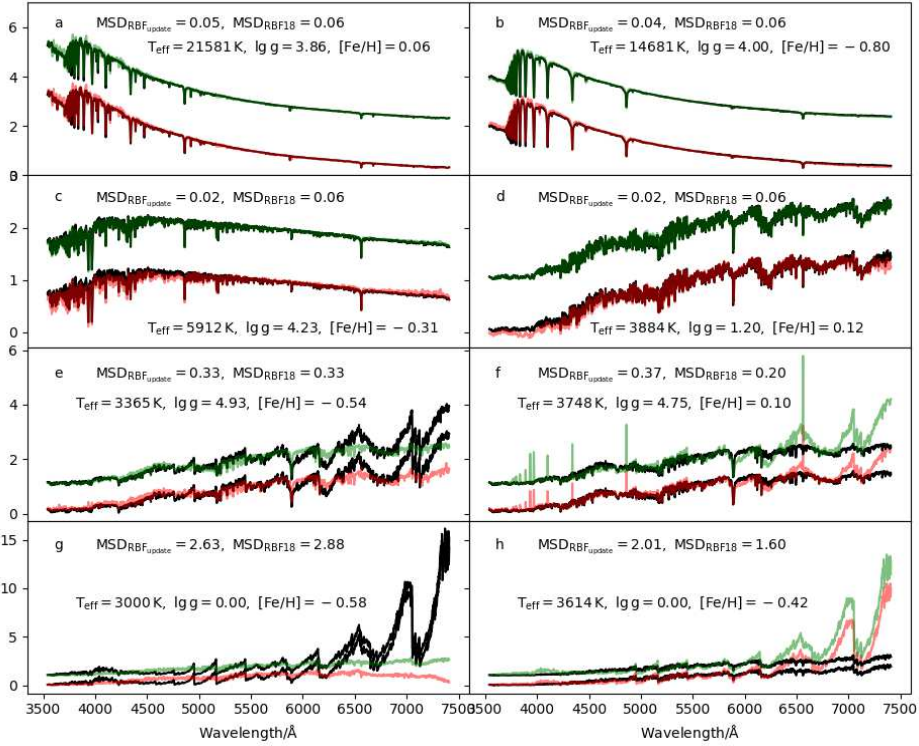}
    \caption{The spectra of eight test stars in the MILES library represented by the red "+" in Fig. \ref{fig: MSE_distribution}. The stellar atmospheric parameters and the test result $\rm MSE$ of $\rm RBF_{update}$ and $\rm RBF18$ interpolators also are listed. For every panel, three spectra are included. The black line is the original spectrum of test stars in MILES library. The green and red translucent lines are the interpolation spectra of test star by $\rm RBF_{update}$ and $\rm RBF18$ interpolators, respectively. For the reason of clarity, the interpolation spectra by $\rm RBF_{update}$ interpolator is shifted upwards, and the corresponding original spectrum also has a copy spectrum moved upwards by the same distance. Moreover, for avoiding overlap, the red and green lines are translucence. First 4 panels (a-d) list four representative test spectra in different areas of stellar atmospheric space. Panels e and f list the representative bad interpolation spectra in the low-mass main sequence region. The last two panels g and h list the representative bad interpolation spectra in red giant branch region.}
   \label{fig: spec_comp}
\end{figure*}	

\begin{figure*}
	\centering
	\includegraphics[width=1\textwidth]{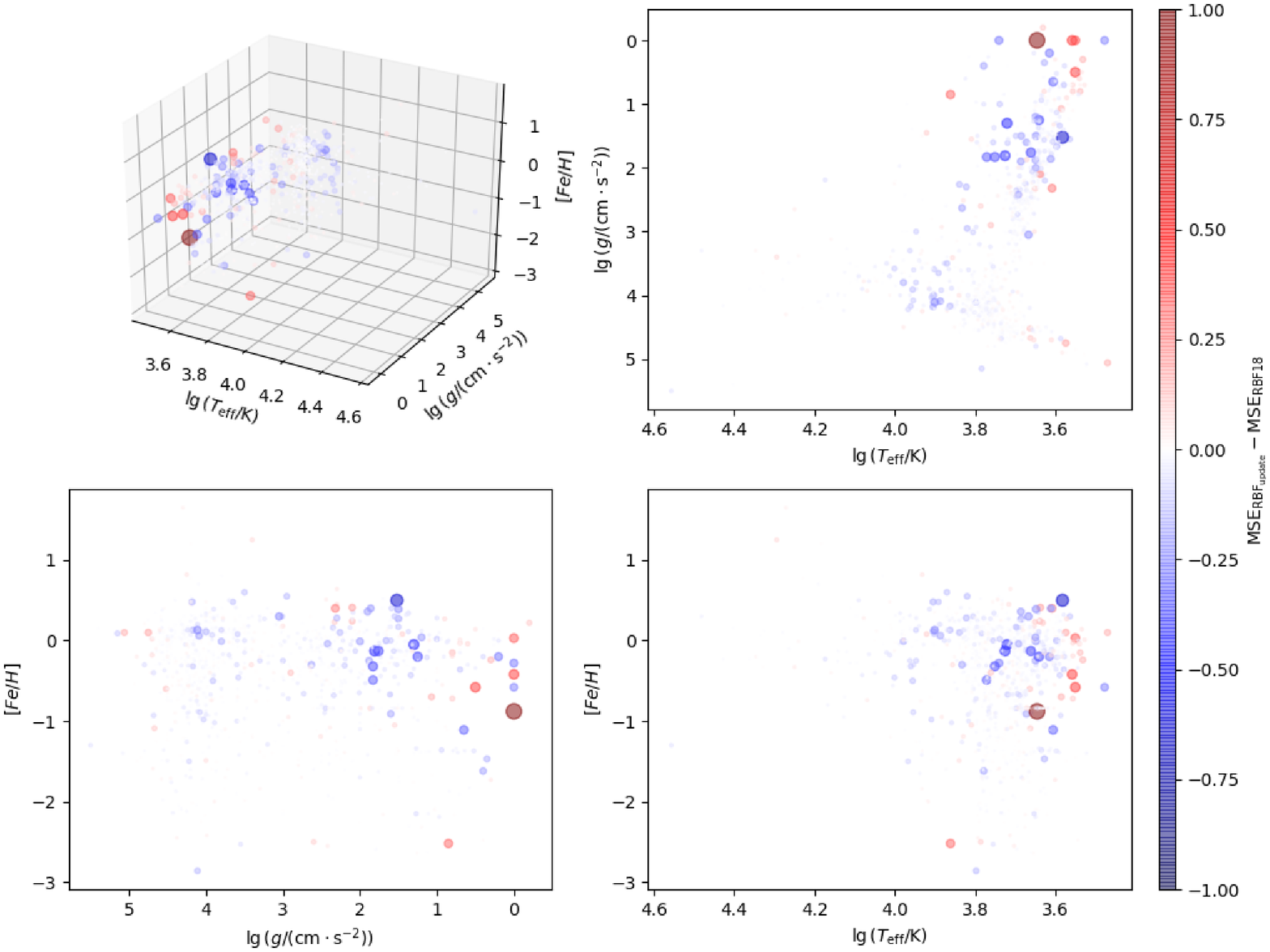}
    \caption{The discrepancy in the $\rm MSE$ between the $\rm RBF_{update}$ and $\rm RBF18$ interpolators based on the MILES library. The points give the positions of test stars. The color of points is used to characterize the value $\rm MSE_{RBF_{update}} - \rm MSE_{RBF18}$, the blue points mean negative value, the red points mean positive value. The point size is used to characterize the absolute value of $\rm MSE_{RBF_{update}} - \rm MSE_{RBF18}$. The top-left panel list the result in 3D stellar atmospheric parameters, the top-right panel gives the projection on the ${\rm lg}\,T_{\rm eff}$ and ${\rm lg}\,g$ plane, the bottom-left panel gives the projection on the ${\rm lg}\,g$ and $[Fe/H]$ plane, the bottom-right panel shows the projection on the ${\rm lg}\,T_{\rm eff}$ and $[Fe/H]$ plane. 
    }
   \label{fig: compare_old}
\end{figure*}

\subsection{Tests and comparison}
	\label{subsec: test_result}
 
In this part, we test $\rm RBF_{update}$ interpolator based the empirical spectral library MILES in Section \ref{subsec: test}, and we give a comparison between the $\rm RBF_{update}$ and $\rm RBF18$ interpolators in Section \ref{subsec: vs_18}. 

\subsubsection{Test of the $RBF_{update}$ interpolator} 
	\label{subsec: test}
	
We test the $\rm RBF_{update}$ interpolator based on the MILES stellar spectral library. In the test, we delete one member star from MILES stellar spectral library and use the remained spectra as a library to calculate the spectrum of the deleted star. Every star in the MILES library has been tested by the above process. And the comparison between interpolated and original spectra is used to show the interpolation performance of $\rm RBF_{update}$ interpolator, (the same test is also done for the spectral interpolator $\rm RBF18$). 

The $\rm MSE$ between interpolated and original spectra is used to characterize the difference,
${\rm MSE} =\sqrt{\sum_{\lambda}( f_{{\rm int},\, \lambda}- f_{{\rm ori}, \, \lambda})^2/{\rm len}(f_{{\rm ori},\, \lambda})}$, where $f_{\rm int,\, \lambda}$ and $f_{\rm{ori},\lambda}$ are the normalized flux on the $\lambda\,$th wavelength interval of the interpolated and original spectra, respectively, ${\rm len}(f_{{\rm ori},\, \lambda})$ is the array length of the spectrum. Every star of MILES library has a corresponding $\rm MSE$ value, we present it in Fig. \ref{fig: MSE_distribution}.

Fig. \ref{fig: MSE_distribution} plots the overall result of the $\rm MSE$ distribution. The top-left panel shows the result in the ${\rm lg}\,T_{\rm eff},\ {\rm lg}\,g,\ [Fe/H]$ space. For the sake of clarity, the top-right panel shows the projection of $\rm MSE$ on the ${\rm lg}\,T_{\rm eff}$ and ${\rm lg}\,g$ plane, the bottom-left panel shows the projection on the ${\rm lg}\,g$ and $[Fe/H]$ plane and the bottom-right panel shows the projection on the ${\rm lg}\,T_{\rm eff}$ and $[Fe/H]$ plane. For each panel, the $\rm MSE$ value is characterized by the gray level and point size. We can find that most stars have relatively small $\rm MSE$ values. The obvious difference exists in parts of the lower-temperature region, especially for stars in the edge of the low-metallicity red giant region. 

For a more detailed analysis of the $\rm RBF_{update}$ interpolator, eight representative stellar spectra are shown in Fig. \ref{fig: spec_comp} (the position of those eight test stars are marked by red "+" in Fig. \ref{fig: MSE_distribution}). 
In each panel of Fig. \ref{fig: spec_comp}, black lines are the original spectrum of the test star, the green and red translucent lines are the interpolation spectra of the test star by $\rm RBF_{update}$ and $\rm RBF18$ interpolators, respectively. The interpolation spectra of the $\rm RBF_{update}$ interpolator and a copy of the original spectrum are moved upwards for the reason of clarity. The stellar atmospheric parameter and the $\rm MSE$ value also are given in each panel.
Here the combination of green and black spectra gives a direct spectrum performance of the $\rm RBF_{update}$ interpolator, the combination of red and black spectra gives a direct spectrum performance of the $\rm RBF18$ interpolator. Panels a-d list the representative spectra of massive main sequence, medium mass main sequence, low-mass main sequence and red giant test stars, we can find the interpolation spectra have a good match with the original spectra. Most of the test stars have similar results in our test, but there are still a few test stars have bad test results.  
Panels e-h show four typically spectra of bad performance by $\rm RBF_{update}$ and $\rm RBF18$ interpolators, they are in the red giant (g-h) and low-mass main sequence (e-f) regions. This bad performance will be discussed in Section \ref{subsec: analysis}.

\subsubsection{Comparison with $\rm RBF18$ interpolator}
	\label{subsec: vs_18}
In this section, we give a comparison between the $\rm RBF_{update}$ and $\rm RBF18$ interpolators by the test in Section \ref{subsec: test}. For any test star in MILES library, we use the $\rm MSE_{RBF_{update}} - \rm MSE_{RBF18}$ to character the discrepancy. The mean square error $\rm MSE$ is large than zero, the smaller value of $\rm MSE$ means the better match between the interpolated and the original spectra of test star in MILES library. So, the negative value of $\rm MSE_{RBF_{update}} - \rm MSE_{RBF18}$ means that the $\rm RBF_{update}$ interpolator has a better performance, the positive value means that the $\rm RBF18$ interpolator has a better performance.

In Fig. \ref{fig: compare_old}, $\rm MSE_{RBF_{update}} - \rm MSE_{RBF18}$ is shown in the stellar atmospheric parameter space. The top-left panel gives the result in $T_{\rm eff}, \ {\rm lg}\, g$ and $[Fe/H]$ space. The other three panels list the corresponding three projections that are same as those in Fig. \ref{fig: MSE_distribution}. The point size is used to characterize the absolute value of $\rm MSE_{RBF_{update}} - \rm MSE_{RBF18}$, the point color is used to characterize the value $\rm MSE_{RBF_{update}} - \rm MSE_{RBF18}$. Here, blue and red points mean negative and positive value of $\rm MSE_{RBF_{update}} - \rm MSE_{RBF18}$, respectively. 

On the whole, the test results of $\rm RBF_{update}$ interpolator are better than those of the $\rm RBF18$ interpolator. In the high temperature main sequence region, it does not have obvious difference for two interpolators. Panels a and b in Fig. \ref{fig: spec_comp} give two typical test results, and both two interpolators have smaller $\rm MSE$.    
In dense part of low-temperature main sequence and red giant regions, the $\rm RBF_{update}$ interpolator has better performance than $\rm RBF18$ interpolator with a lower $\rm MSE$ value. Two typical test spectra are shown in panels c and d in Fig. \ref{fig: spec_comp}, the green interpolation spectra have a better match with the original spectra than the red one.

But for some test stars on the edge of the low-temperature region, both $\rm RBF_{update}$ and $\rm RBF18$ interpolators have bad performance, the spectra of four typical test stars in those regions are shown in panels e-h of Fig. \ref{fig: spec_comp}, the detail analysis of the reason will be given in Section \ref{subsec: analysis}.

\begin{figure}
	\centering
	\includegraphics[scale=0.4]{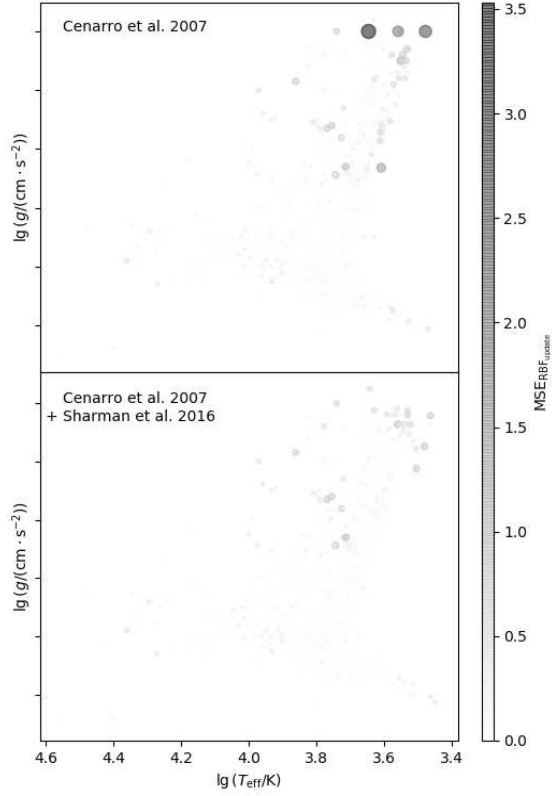}
    \caption{The $\rm MSE$ distribution of the test for the $\rm RBF_{update}$ interpolator on the ${\rm lg}\,T_{\rm eff}$ and ${\rm lg}\,g$ plane. The top panel is the same as the top-right panel of Fig. \ref{fig: MSE_distribution} and gives the result $\rm MSE$ distribution of the test which is based on the MILES library with its original stellar atmospheric parameters \citep{2007MNRAS.374..664C}. In the bottom panel, we show the $\rm MSE$ distribution of test based on the MILES library with the corrected cool stellar atmospheric parameters \citep{2016A&A...585A..64S}. For every panel, the grey level and the size of points are used to characterize the $\rm MSE$ values. 
    }
   \label{fig: test_mix}
\end{figure}

\subsection{Analysis of result}
	\label{subsec: analysis}
Here, we will give an analysis of the test result in Section \ref{subsec: test_result}.
 On the whole, the $\rm RBF_{update}$ interpolator has good performance than the $\rm RBF18$ interpolator in the spectral interpolation calculation. However, it has bad performance in the edge of the low-temperature region.
A typical example of bad performance is shown in the last four panels of Fig. \ref{fig: spec_comp}. Panels e and f show the test result of two test stars in the edge of the low-temperature main sequence region, panels g and h show the test result of two test stars in the edge of the red giant region. The positions of these four test stars are shown in Fig. \ref{fig: MSE_distribution}, we can find the test stars of panels e and f are adjacent and those of panels g and h are adjacent. 
From the Gaussian kernel function (Eq. \ref{eq: interpolator}), we know that the spectrum of the adjacent star has a bigger effect than the distant one in the interpolation calculation. Therefore, in the test, the calculation of the deleted spectrum depends largely on the adjacent spectra in the stellar atmospheric parameter space. 
For the test star in panel e, the star f has a big effect, that is the reason the interpolation spectrum in panel e is similar to the original spectrum in panel f. For the interpolation spectrum of panels f, g and h, the situations are similar.

For the empirical stellar spectral library, we list three possible reasons for the bad test results. 
\begin{enumerate}   
	\item The finite spectra faces its complex change in some stellar atmospheric parameter region. It means that the library is incomplete and has not include enough typical spectra.
	\item Three stellar atmospheric parameters can not determine solely the spectra. It means that one set of stellar atmospheric parameters in 3-D space corresponds to several potential spectra with obvious difference\footnote{An example is that $[Fe/H]$ can not describe the ratio of the different elements in the stellar atmosphere, this problem has not an obvious effect in the high-temperature region, but can not be ignored in the low-temperature region.} (similar to the description in \citealt{2019A&A...627A.138A}.) 
	\item The stellar atmospheric parameters are not self-consistent. The stellar spectra do not vary with the stellar atmospheric parameters by a one-by-one relation\footnote{A simple example is that a smooth change of the spectra in the stellar atmospheric parameter space can be broken and becomes messy by adding a set of random biases on the stellar atmospheric parameters in the library.}.  
	
\end{enumerate}

For the first reason, more targeted observational data are needed. For the second reason, more potential parameters of spectra should be given for a more strict constraint on the spectra. For the third reason, the spectra in the library need a more detailed derivation of atmospheric parameters, we give a test by using those self-consistent stellar atmospheric parameters. 
Here we use a relatively new result of \citet{2016A&A...585A..64S} to test this idea. 
In \citet{2016A&A...585A..64S}, $\sim 300$ cool stars in MILES library were refined. We replace the corresponding parameter of MILES library by the results of \citet{2016A&A...585A..64S}, and use them to test $\rm RBF_{update}$ interpolator as did in Section \ref{subsec: test}. In Fig. \ref{fig: test_mix}, we show the test results. The top panel shows the $\rm MSE$ distribution of test stars based on the MILES library with the original stellar atmospheric parameters \citep{2007MNRAS.374..664C}, it is same as the top-right panel in Fig. \ref{fig: MSE_distribution}, the bottom panel shows the MES distribution of the test based on the MILES library with replaced stellar atmospheric parameters of the cool stars \citep{2016A&A...585A..64S}. An obvious improvement is appeared in the bottom panel, this result prove the self-consistent parameters of library is important for the spectral interpolation calculation.

\section{Conclusions}
	\label{sec: conclusion}
In stellar population synthesis models, the empirical stellar spectral library is necessary for the integrated spectra of the stellar populations. 
In this work, we improve the RBF network by comparing with the other kernel methods (SPHs and likelihood approximation) and give an upgraded spectral RBF interpolator. We include a constraint about the kernel function (Eq. \ref{eq: dens-sigma}) in the RBF network, this constraint gives the relation between the $\sigma$ of Gaussian kernel function and the sample spatial density in the parameter space. 

Moreover, we also consider the anisotropic kernel function by relating it to the inhomogeneous distribution of stars in the stellar atmospheric parameter space. We use the local axial direction dispersion to determine the anisotropic kernel function (Eq. \ref{eq: local dispersion}).
By including three control parameters $c_0, c_1, c_2$, we can get a RBF network for spectral interpolation calculation, here we call it $\rm RBF_{update}$ interpolator. The BAS search algorithm is used to search the best control parameters $c_0, c_1, c_2$ by matching with the semi-empirical BaSeL-3.1 stellar spectral library.
 
We also use a test to analyze the performance of $\rm RBF_{update}$ interpolator based on the MILES stellar spectral library. In the test, we select any star in the MILES library as the test object and compare the original with the interpolation spectra which is calculated by the $\rm RBF_{update}$ interpolator based on the remained stellar spectra in MILES library. 
We find that $\rm RBF_{update}$ interpolator has a good performance in general except for some test stars in the edge of the red giant and low-temperature main sequence regions (Fig. \ref{fig: MSE_distribution}). 

Three possible reasons can cause these bad performance for empirical stellar spectral library, the first is the incomplete spectral coverage in the stellar atmospheric parameter space, the second is the existence of potential atmospheric parameters, the third is the inconsistent atmospheric parameters. For the first two reasons, more observations are needed and the additional atmospheric parameters should be included in the stellar spectral library. 
For the last reason, the modified stellar atmospheric parameters of the stellar spectral library are needed. 
Moreover, we also give a comparison between the $\rm RBF_{update}$ interpolator and our early work in \citet{2018MNRAS.476.4071C}, the results show that the  $\rm RBF_{update}$ interpolator has an obvious improvement, except for the edge of the low-temperature region.  (Fig. \ref{fig: compare_old}), the same reasons make both interpolators have not a good performance in these regions.

At last, the code of $\rm RBF_{update}$ interpolator is written by Python and you can find it in \url{http://www1.ynao.ac.cn/~zhangfh/}. The code can be used for different libraries and user can use it with the modified stellar spectral library by adding additional spectra or updating the stellar atmospheric parameters of the library.

\normalem
\begin{acknowledgements}
This project was partly supported by the Chinese Natural Science Foundation (No. 11973081 and 11521303), the Yunnan Foundation (grant No. 2011CI053 and 2019FB006) and the Youth Project of Western Light of Chinese Academy of Sciences. We also thank the referee for suggestions that have improved the quality of this manuscript.
\end{acknowledgements}
  
\bibliographystyle{raa}
\bibliography{paper}

\end{document}